# VERIFYING THE RADIOSONDE HUMIDITY SENSOR PERFORMANCE

Kochin Alexander

Central Aerological Observatory, Pervomaiskaya st. 3, Dolgoprudny, Moscow Region, Russia, 141701

Corresponding author: Alexander Kochin, *amarl@mail.ru*

**Abstract:**

Monitoring the performance of humidity sensors has become particularly relevant due to the lack of in-house production of humidity sensors in the Russian Federation and the logistical problems that have arisen. Humidity is subject to high spatial variability, therefore, standard methods for monitoring data quality based on the difference with the field of the first approximation are not applicable for its control. It is proposed to carry out monitoring by comparing readings on the surface and at altitude with a pressure of 850 hPa, where humidity is less than 60% in the absence of low clouds, and more than 70% in its presence. The fact of the presence/absence of low clouds is determined by the readings of a vertically oriented pyrometer. The difference between the air temperature at the surface and the temperature from the pyrometer of more than 14°K corresponds to the absence of low clouds, and less than 14°K corresponds to the presence of low clouds. Accordingly, a serviceable humidity sensor should show the appropriate values.

## 1. Introduction

Monitoring the quality of observational data is necessary to ensure the reliability of the information received. The temperature and wind speed are characterized by relatively low spatial variability, which makes it possible to create a quality monitoring system. Examples of comparisons of measurement data with the first approximation field using the national monitoring system of the Russian Federation are shown in Fig.1.

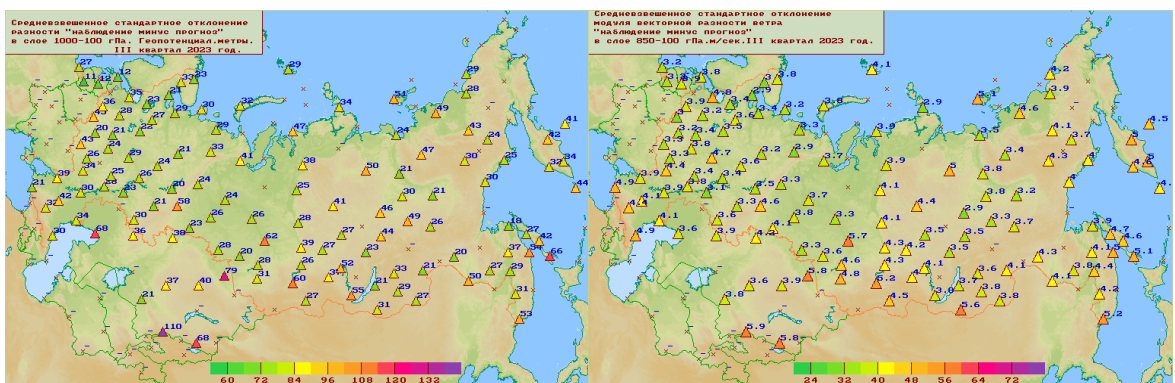

**Fig.1. Results of data quality monitoring for the third quarter of 2023. Temperature data on the left, wind direction on the right. Upper-air sounding stations with unsatisfactory data quality are marked in red.**

Humidity is characterized by high spatial variability, therefore, comparison with the field of the first approximation does not provide quality control of humidity sensors. The manufacturer checks the sensors under normal conditions, but this does not guarantee their performance at low pressure and low temperatures. The coincidence of one reading with ground data before launch does not guarantee the correctness of measurements. A second reference point at a certain height provides some confidence in the quality of measurements. Errors in humidity measurements occur not only due to poor-quality sensors, but also due to staff errors. For example, common mistakes are not removing the transport protective cover or incorrect mounting of the sensor when assembling the radiosonde. In Europe and America, the problem is solved by comparing it with the data of the AMDAR aircraft system. This aircraft system is not used in the Russian Federation. Therefore, the purpose of the work is to prepare for the creation of an automatic error warning system. Humidity profiles in clouds and in a cloudless atmosphere have specific features that can be used to check humidity sensors if the presence/absence of clouds is reliably determined.

## 2. The method of checking the humidity sensor

Air temperature and wind direction are conservative characteristics, the average values of which change slowly in space and time. The radius of spatial correlation is determined by the size of mesoscale processes such as cyclones and anticyclones. The distance between upper-air sounding stations is determined precisely by a similar synoptic scale of about 200-300 km. To monitor the measurement quality of air temperature and wind direction, a comparison method with the field of the first approximation is used. Humidity data does not pass a similar test. Perhaps the implementation of such a procedure could clarify the quality of humidity measurements, but this task has not yet been solved.

Checking the sensor's operability requires comparing its readings with a certain reference value at least at two points. The first control point is the comparative humidity measurements before release at the pre-flight control. As a second reference point, the humidity value at a surface height of 850 hPa can be used. The peculiarity of humidity at an altitude of 850 hPa is its close connection with the presence of low clouds. In the presence of low clouds, the humidity will be more than 70%, and in the absence of less than 60%. It is assumed that the presence of clouds will be detected using a cloud detector based on a pyrometer (Lui Liu 2015, WMO 2023) in automatic mode. Humidity in the surface layer in the absence of clouds can take any values and cannot be used for control.

## 3. The results of the observation

Examples of radiosonde data at the Dolgoprudnaya upper-air sounding station (27713) are shown in Table 1.

**Table 1: Examples of radiosonde data**

| Time | REL H % at the surface | REL H % at 850 hPa | HGHT 850 hPa m | Cloudiness |
|---|---|---|---|---|
| 28.11.23.00Z | 77 | 83 | 1006 | Y |

| 28.11.23.12Z | 77 | 77 | 1 095 | **Y** |
| --- | --- | --- | --- | --- |
| 18.11.23.00Z | 74 | 28 | 1265 | **N** |

Thus, the fact of the presence/absence of clouds forms a second control point at an altitude of 850 HPa.

The presence or absence of clouds is determined by the pyrometer data and by the magnitude of the surface temperature. The difference between the air temperature at the surface and the temperature from the pyrometer of more than 14 °K corresponds to the absence of clouds, and less than 14°K corresponds to the presence of low clouds (WMO 2023). The algorithm was tested at the Dolgoprudnaya upper-air sounding station (27713), where a pyrometer was installed. To control the presence of clouds and determine the cloud base height, the observer recorded the moment when the balloon disappeared into the cloud. Using this technique, 10 launches were carried out, which confirmed the correctness of the proposed algorithm for determining the presence of clouds by pyrometer.

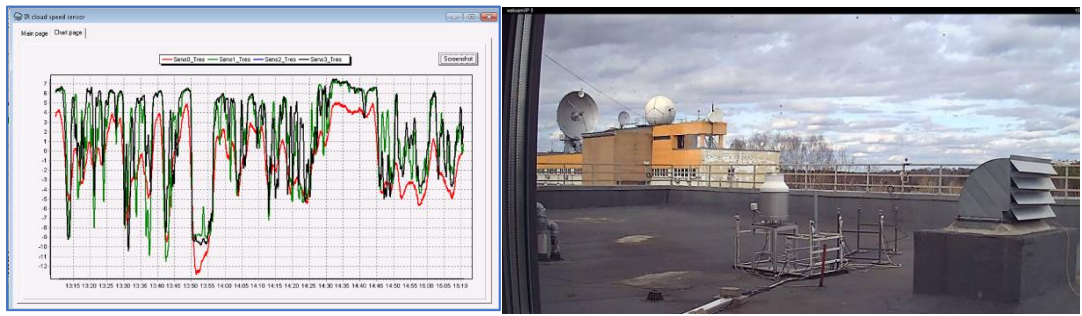

**Fig. 2. An example of pyrometer temperature in the presence of clouds, as in the photo on the right.**

### 4. Conclusions

Satellite methods measure water vapor concentration profiles remotely. However, the procedure for restoring the vertical humidity profile according to satellite profiler data requires solving an incorrect inverse problem. To solve an incorrect inverse problem, a specific type of profile is required, which uses upper-air sounding data. Therefore, the accuracy of the humidity measurement by the radiosonde determines the final accuracy of the restoration of the humidity profile by the satellite profiler. Checking the sensor's operability requires comparing its readings with a certain reference value at least at two points. The first control point is comparative humidity measurements before release at pre-flight control. As a second reference point, it is proposed to use the humidity value at a surface height of 850 hPa. Humidity at this altitude is more than 70% if there is low cloud cover, and less than 60% if there is no cloud cover. The presence or absence of clouds is determined by the pyrometer data and by the magnitude of the surface temperature. The difference between the air temperature at the surface and the temperature from the pyrometer of more than 14°K corresponds to the absence of clouds to a height of at least 1,500 meters, and less than 14°K corresponds to the presence of low clouds (WMO 2023). The algorithm was tested at the Dolgoprudnaya upper-air sounding station (27713), where a pyrometer was installed. As an additional point for monitoring the humidity sensor, we can use the effect of a sharp decrease in humidity at the upper boundary of the cloud. The nature of humidity changes according to data from different radiosondes is shown in Fig.3 (Kochin & al. 2021.)

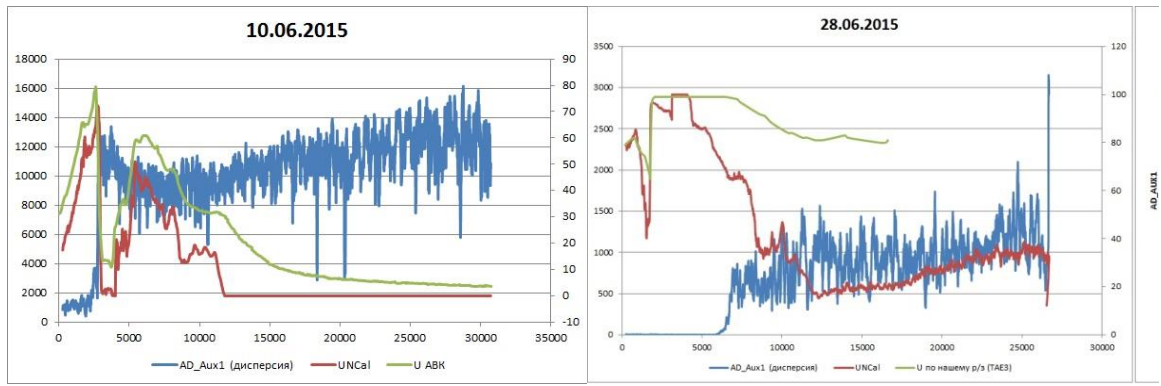

**Fig.3 Cloud without precipitation on the left (red line humidity sensor MODEM M2K2C, green – MRZ-3). At the exit from the cloud, both sensors synchronously showed a sharp change in humidity. Above the tropopause, the MRZ-3 sensor significantly overestimates humidity. On the right there are powerful cumulonimbus cloud with precipitation. The M2K2C sensor "dried up" quickly enough, the MRZ-3 sensor did not return to its original state.**

**Acknowledgments.**

The author thanks the staff of the Dolgoprudny upper-air sounding station for their assistance in radiosonde launches.